\documentclass[preprint2]{aastex6}

\usepackage{natbib}

\begin{document}

\title{Simultaneous VLBI astrometry of H$_{2}$O and SiO Masers toward the Semiregular Variable R Crateris}

\author{Dong-Jin Kim\altaffilmark{1,2,3}}
\author{Se--Hyung Cho\altaffilmark{2,1}}
\author{Youngjoo Yun\altaffilmark{2}}
\author{Yoon Kyung Choi\altaffilmark{3}}
\author{Dong--Hwan Yoon\altaffilmark{2,5}}
\author{Jaeheon Kim\altaffilmark{4}}
\author{Richard Dodson\altaffilmark{6}}
\author{Mar\'{\i}a J. {Rioja}\altaffilmark{6,7,8}}
\author{Haneul Yang\altaffilmark{2,5}}
\author{Suk--Jin Yoon\altaffilmark{1}}

\altaffiltext{1}{Department of Astronomy, Yonsei University, 50 Yonsei-ro Seodaemun-gu, Seoul, 03722, Republic of Korea}
\altaffiltext{2}{Korea Astronomy and Space Science Institute, 776 Daedeok--daero, Yuseong--gu, Daejeon 34055, Korea; cho@kasi.re.kr}
\altaffiltext{3}{Max-Planck-Institut f$\ddot{\textrm u}$r Radioastronomie, Auf dem H$\ddot{\textrm u}$gel 69, 53121 Bonn, Germany}
\altaffiltext{4}{Shanghai Astronomical Observatory, Chinese Academy of Sciences, Shanghai 200030, China}
\altaffiltext{5}{Astronomy program, Department of Physics and Astronomy, Seoul National University, 1 Gwanak{ro, Gwanak{gu, Seoul 08826, Korea}}}
\altaffiltext{6}{International Center for Radio Astronomy Research, M468, The University of Western Australia, 35 Stirling Hwy, Crawley, Western
Australia, 6009, Australia}

\altaffiltext{7}{Observatorio Astron\'omico Nacional (IGN), Alfonso XII, 3 y 5, 28014 Madrid, Spain}
\altaffiltext{8}{CSIRO Astronomy and Space Science, 26 Dick Perry Avenue, Kensington WA 6151, Australia}

\begin{abstract}

We obtained, for the first time, astrometrically registered maps of the 22.2 GHz H$_{2}$O and 42.8, 43.1 and 86.2 GHz SiO maser emission toward the semiregular b-type variable (SRb) R Crateris, at three epochs (May 21st, 2015, Jan 7th and Jan 26th, 2016) using the Korean VLBI Network. The SiO masers show a ring-like spatial structure, while the H$_{2}$O maser shows a very asymmetric one-side outflow structure, which is located at the southern part of the ring-like SiO maser feature. We also found that the 86.2 GHz SiO maser spots are distributed in an inner region, compared to those of the 43.1 GHz SiO maser, which is different from all previously known distributions of the 86.2 GHz SiO masers in variable stars. The different distribution of the 86.2 GHz SiO maser seems to be related to the complex dynamics caused by the overtone pulsation mode of the SRb R Crateris. Furthermore, we estimated the position of the central star based on the ring fitting of the SiO masers, which is essential for interpreting the morphology and kinematics of a circumstellar envelope. The estimated stellar coordinate corresponds well to the position measured by {\sc gaia}.

\end{abstract}


\section{Introduction}

It is well known that Asymptotic Giant Branch (AGB) stars rapidly lose their mass due to the strong stellar wind. The dense and strong stellar wind forms a circumstellar envelope (CSE), which is the main site of SiO, H$_{2}$O, and OH masers around oxygen-rich AGB stars. Various SiO maser lines arise from a distance of 2--4 stellar radii and show a ring-like structure with inflow or outflow motions below a dust formation layer \citep{1994ApJ...430L..61D,2003ApJ...599.1372D,2015A&amp;A...576A..70P}. The 22.2 GHz H$_{2}$O maser arises partially in the dust formation layer and partially at greater radii above the dust layer and represents acceleration motions of stellar winds in the CSE \citep{1978ApJ...222..132R,1981ARA&amp;A..19..231R,2003ApJ...590..460I}. Maps of the H$_{2}$O and SiO masers, registered to the stellar continuum, indicate that the star is at the center of shell-like maser emission in the case of W Hya \citep{1990ApJ...360L..51R,2007ApJ...671.2068R}. Therefore, a combined study of the H$_{2}$O and SiO masers enables us to investigate the formation and development of stellar winds. 

However, previous VLBI observations of the H$_{2}$O and SiO masers have been performed separately, due to the lack of a simultaneous observation system for the H$_{2}$O and SiO masers. The Korean VLBI Network (KVN) is equipped with a quasi-optics system for simultaneous observations at the K (21--23 GHz), Q (42--44 GHz), W (85--95 GHz), and D (125--142 GHz) bands \citep{2008IJIMW..29...69H}. Therefore, the KVN Key Science Project (KSP) on evolved stars was started for the combined study of H$_{2}$O and SiO masers in the CSE (\url{https://radio.kasi.re.kr/kvn/ksp.php}, \citet{2018IAUS..336..359C}). The first stage of the KSP on evolved stars focused on nine objects for which astrometrically registered maps for both the H$_{2}$O and SiO masers have been successfully obtained.

Here we present the astrometrically registered maps for both the H$_{2}$O and SiO maser lines at three epochs for the semiregular variable star R Crateris (R Crt). R Crt is classified as a SRb star with the spectral type of M7 \citep{2017ARep...61...80S}. SRc type indicates super-giant stars, and SRa type variables show a persistent periodicity with smaller light-amplitudes ($<$2.5 mag in V) than Mira variables. In contrast, SRb type variables show an uncertain or superimposed periodicities, such as one or more overtones. The role of the overtone pulsation mode on the maser properties and the mass loss process has rarely been investigated. In this sense, R Crt is a high priority target for the study of overtone-pulsators emitting SiO, H$_{2}$O and OH masers \citep{2001A&amp;A...378..522E, 2010ApJS..188..209K}. The mass loss rate of R Crt was estimated to be 8.0$\times$10$^{-7}$M$_{\odot}$yr$^{-1}$ \citep{2002A&A...391.1053O}. It has a relatively higher mass loss rate than that of the usual semiregular variables \citep{1998ApJS..117..209K}. The estimated distance to R Crt is some what ambiguous, with estimates ranging from 170 to 300 pc \citep{1998ApJS..117..209K, 1999MNRAS.304..415S, 2001PASJ...53.1231I}.

\section{Observations and Data Reduction}

We performed simultaneous VLBI monitoring observations of H$_{2}$O 6$_{16}$--5$_{23}$ (22.23508 GHz) and SiO v=1, 2, J=1$\rightarrow$0, SiO v=1, J=2$\rightarrow$1, 3$\rightarrow$2 (43.12208, 42.82058, 86.24344, 129.36335 GHz) masers toward R Crt with the KVN, which consists of three 21 m radio telescopes \citep{2011PASP..123.1398L}. The monitoring was carried out at 11 epochs from Oct 2014 to Feb 2016. In this paper, we present three epochs of observations, which show the astrometrically registered maps of the H$_{2}$O and SiO masers. The remaining data will be presented in a forthcoming paper (D. J. Kim et al. in prep.). The correlator coordinates used for R Crt were R.A.=11:00:33.850,  Dec.=--18:19:29.60. The size of the synthesized beams are typically 6/3/1.5 mas at the K/Q/W bands, and the system temperatures were up to 220/210/450/800 K (epoch 1), 200/300/800/1200 K (epoch 2), and 140/200/400/300 K (epoch 3) at K/Q/W/D bands respectively. In total 16 Intermediate Frequencies (IFs) were used (6/6/2/2 for K/Q/W/D bands), and each IF has a 16 MHz bandwidth.

The schedule consisted of alternating $\sim$2 min scans between the target source R Crt and a continuum calibrator source J1048-1909 using simultaneous 4-band observations. The angular separation between R Crt and J1048-1909 is 3.06 degrees.  J1048-1909 has a positional accuracy of 0.06 mas in R.A. and 0.09 mas in Dec. \citep{2009ITN....35....1M}. A fringe finder, 4C39.25, was also observed for 5 min every hour. Total observation time was about 7 hr for each epoch. We used the Mark5B system for data recording and playback, which has a maximum recording rate of 1 Gbps. The correlation was performed using the DiFX software correlator with a spectral resolution of 512 channels per IF, providing velocity resolutions of 0.42, 0.22, 0.11, and 0.07 km s$^{-1}$ for line observations at K, Q, W, and D bands respectively. We used the Astronomical Image Processing System (AIPS) package for the data reductions. 

We applied conventional phase referencing (PR) techniques for the astrometric measurement of the position of the 22.2 GHz H$_{2}$O maser with respect to the external reference source, J1048-1909. Next, we used the Source Frequency Phase Referencing (SFPR) technique to attain a bonafide astrometric registration of the multiple SiO maser lines (42.8, 43.1, and 86.2 GHz). The combination of the simultaneous multi-frequency capability of the KVN and the SFPR analysis results in precise absolute positions of the maser lines, when the external reference source has precise absolute coordinates. The basis of the SFPR calibration strategy is presented in \cite{2011AJ....141..114R}, and the first application to a spectral line is presented in \cite{2014AJ....148...97D}. The positions of the maser spots in the PRed and SFPRed maps were measured using the two-dimensional Gaussian fitting task in AIPS, and artificial components were filtered out by the requirement that they must appear in more than three successive velocity channels in the same maser feature.

\section{Results} 
As shown in Figure 1, the H$_{2}$O masers toward R Crt are only distributed in the southern part of the SiO masers and spread out over 100$\times$80 mas with several distinct maser features. The spatial distribution of the H$_{2}$O maser is asymmetric, but SiO masers show a ring-like structure of 30 mas in size. The 129.3 GHz SiO maser map was not obtained. Figure 2 displays the total power (solid line) and correlated flux (dotted line) spectra of each maser lines at three epochs. The fraction of missing flux were obtained from single-dish observations using the KVN Yonsei telescope, and they range from 20 to 80\%. As a general trend, the missing flux increases with resolution. The 86.2 GHz SiO maser shows a higher missing flux rate (up to 80 \%) and a wider velocity width than the other maser lines. The full velocity widths of the H$_{2}$O and SiO maser spectra are comparable (from V$_{LSR}$=3 to 16 km s$^{-1}$) except for the 86.2 GHz SiO maser (from V$_{LSR}$=-2 to 18 km s$^{-1}$). The peak flux of the H$_{2}$O maser ($\sim$hundreds of Jy) is much higher than those of the SiO masers ($\sim$dozens of Jy).

\begin{figure}
\figurenum{1}
\includegraphics[scale=0.45]{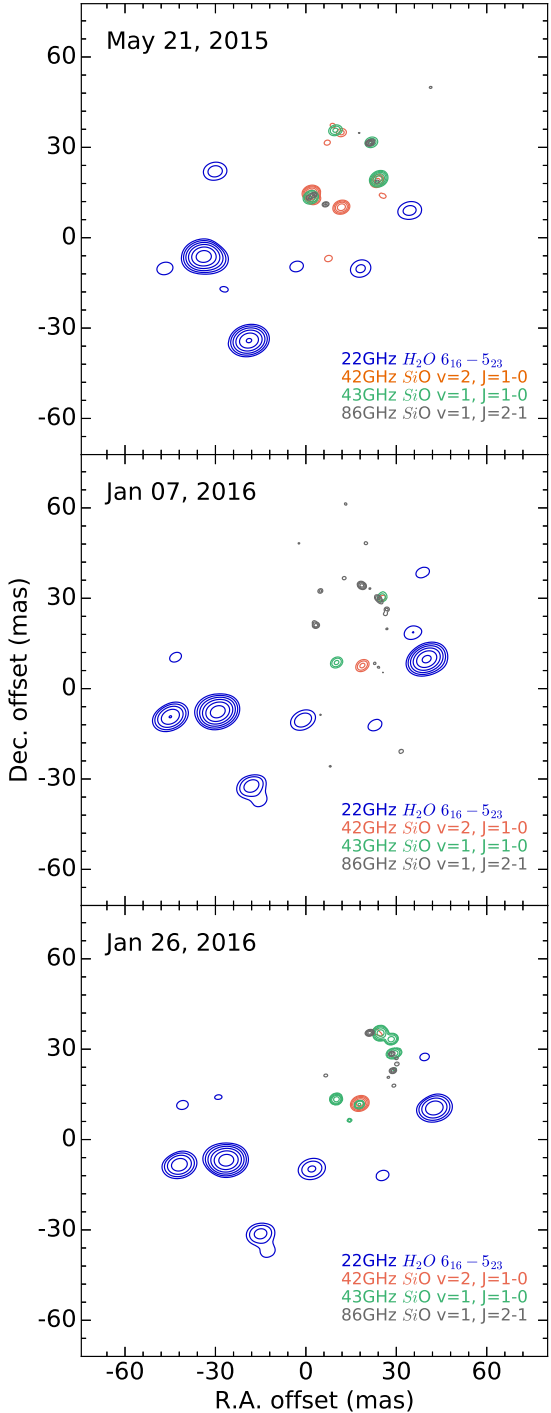}

\caption{Astrometrically registered integrated intensity-velocity maps of the H$_{2}$O and SiO masers.
The peak fluxes of the 22.2/42.8/43.1/86.2 GHz masers are 684.7/7.9/19.5/13.5, 395.5/28.8/2.8/2.8 and 453.8/15.0/1.8/16.0 Jy beam$^{-1}$ km s$^{-1}$ at epoch 1, 2, and 3 respectively. The contour levels are plotted with log scale based on the peak fluxes.}

\end{figure}

\begin{figure*}[ht!]
\centering
\figurenum{2}
\includegraphics[scale=0.4]{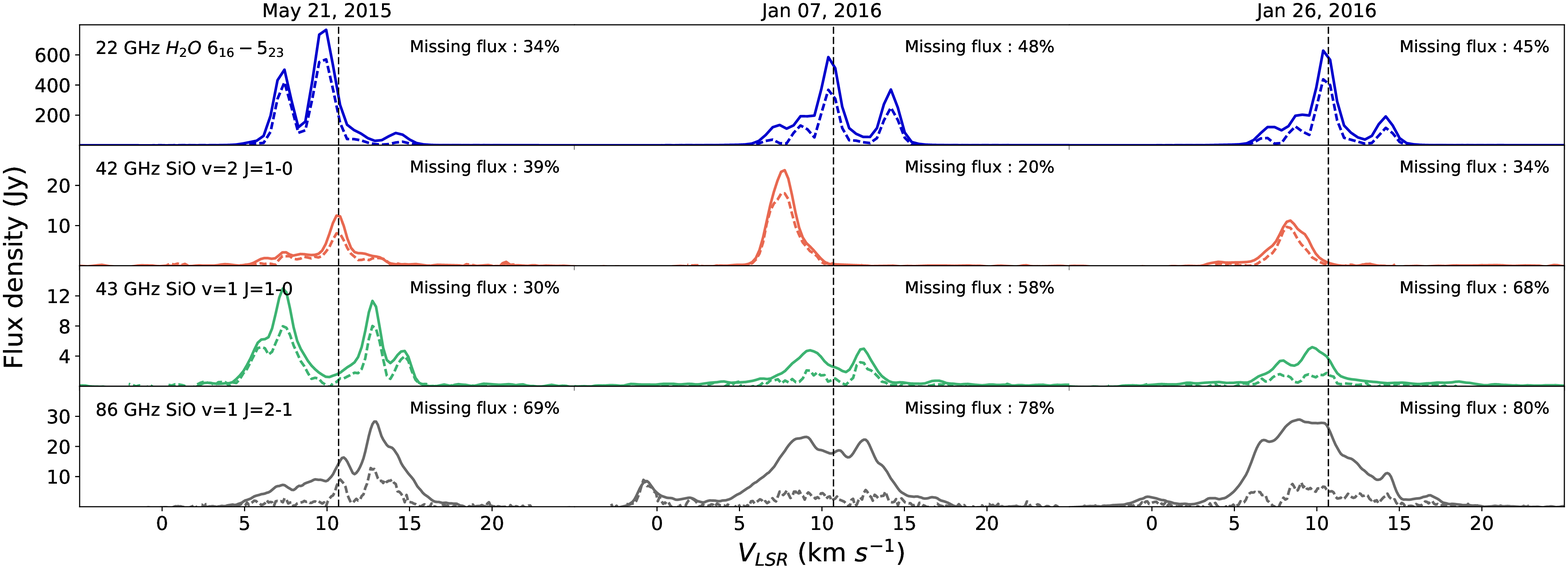}
\caption{Total power (solid) and correlated flux (dashed) spectra of the H$_{2}$O and SiO masers at the 3 epochs of observations. The total power spectra were obtained from the KVN Yonsei telescope. The fraction of missing flux is marked at the top-left corner. Vertical dotted lines indicate the stellar velocity of R Crt (V$_{LSR}$=10.8 km s$^{-1}$)}. 
\end{figure*}

Figure 3 shows the position-velocity spot maps of the masers. The majority of maser spots are blue-shifted in both the H$_{2}$O and SiO masers. The high velocity spots of the 86.2 GHz maser appear at epoch 2 (Jan 7, 2016), and their locations are marked with a blue dashed square in Figure 3. The high velocity spots are blue shifted up to 13 km s$^{-1}$ with respect to the stellar velocity of R Crt (V$_{LSR}$=10.8 km s$^{-1}$), which was measured from the CO J=1$\rightarrow$0 and J=2$\rightarrow$1 lines \citep{1994A&A...290..183K}, and they exceed the terminal velocity of R Crt, 10.3 km $s^{-1}$ \citep{1992A&AS...93..121N}. We estimated the central position and size of the ring-like structure of the SiO masers with a least-square minimization fitting method. There was no fit in the case of the 42.8 GHz SiO maser spots due to their insufficient number. Both 43.1 and 86.2 GHz SiO maser spots were used for determining the central star position. Figure 4 and Table 1 show the fitting results. The deduced absolute coordinate of the central star is RA=11:00:33.8201, Dec.=--18:19:29.618, which is the mean value of epochs 2 and epoch 3. Epoch 1 was excluded in the averaging to minimize uncertain factors such as proper motion and annual parallax. The fitted radius of the ring-like structures is in the range of 13.35 to 13.84 mas for the 43.1 GHz SiO maser and 11.76 to 12.72 mas for the 86.2 GHz SiO maser.
\begin{figure*}[ht!]
\figurenum{3}
\centering
\includegraphics[scale=0.4]{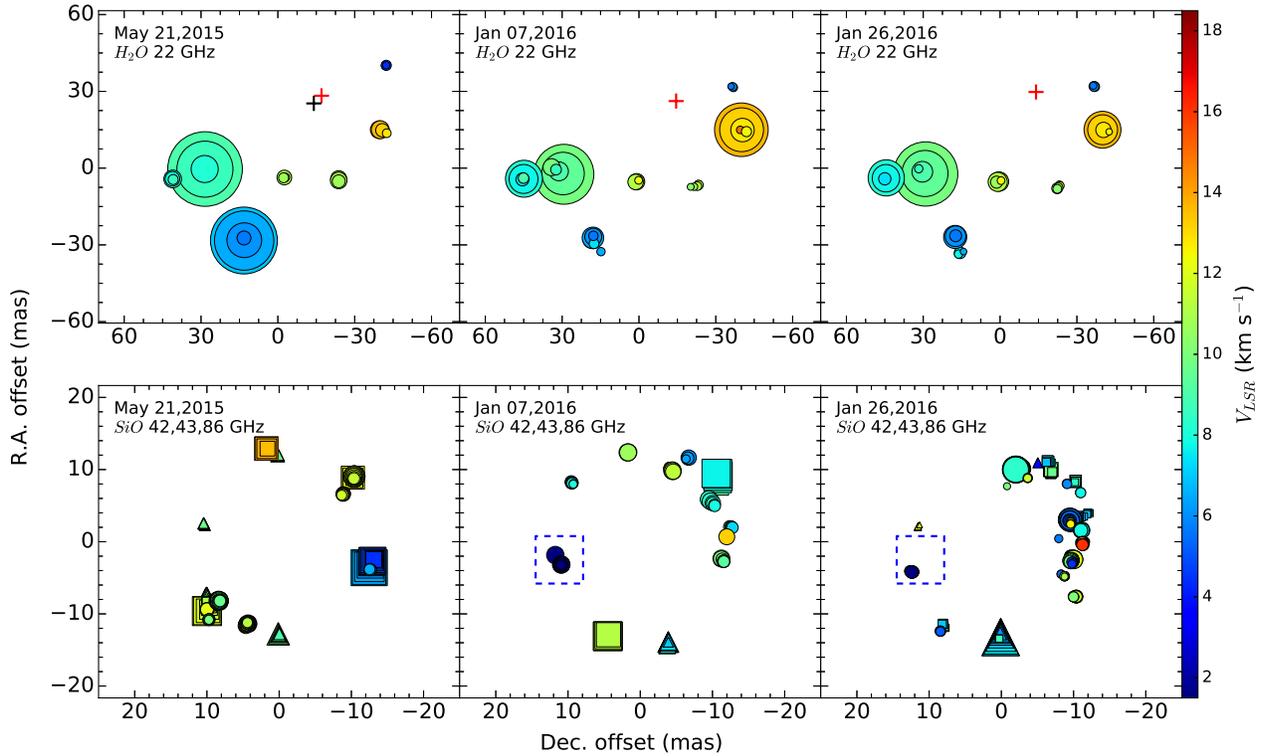}
\caption{The position-velocity spot maps of the H$_{2}$O and SiO masers. The size of a spot is proportional to its flux density. Upper: Maps of the H$_{2}$O maser spots. The red crosses indicate the central positions of the SiO ring-like structures. The black cross represents the position of R Crt measured by {\sc gaia} satellite. Lower: Maps of the 42.8 (triangle), 43.1 (square) and 86.2 (circle) GHz SiO maser spots. The spots inside the dotted-blue squares present the highly blue-shifted components of the 86.2 GHz SiO masers (--2 to 0 km s$^{-1}$).}

\end{figure*}
The H$_{2}$O maser features show coincident spatial distributions over three epochs although they show remarkable changes in their intensities. In the total power spectra, the peak intensity of the blue-shifted components diminish, whilst the red-shifted components increase (Figure 2). The SiO masers present rapid changes in the number of spots, intensities, and positions during a short time interval between epochs 2 and 3 (19 days).

\begin{figure*}[ht!]
\centering
\figurenum{4}
\includegraphics[scale=0.3]{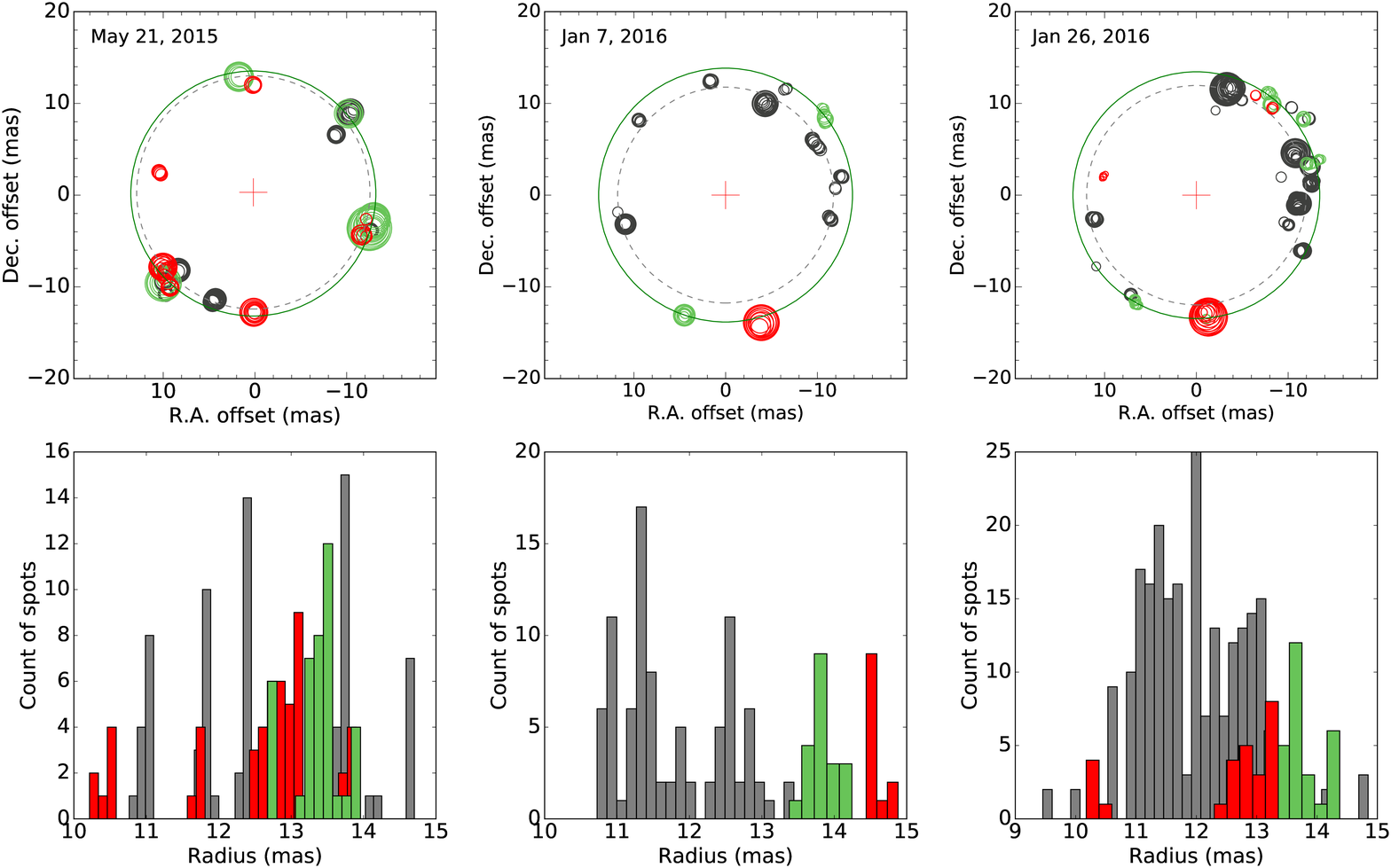}

\caption{Upper: Spatial distributions of the 42.8 (red), 43.1 (green) and 86.2 GHz SiO (grey) maser spots along with superimposed fitted rings for the 43.1 and 86.3 GHz SiO maser spots, shown by the large green continuous and grey-dashed circles, respectively \citep{2018IAUS..336..359C}. The size of an individual spot is proportional to the flux density. The red crosses mark the central position of the rings. Lower: The number of maser spots according to the distance from the center. Each color indicates the 42.8 (red), 43.1 (green) and 86.2 GHz 
SiO (grey) masers.}

\end{figure*}

\floattable
\begin{deluxetable}{cccccccccc}
\tablecaption{The radius and central position of the ring-like structure of the SiO maser lines.}
\tablecolumns{11}
\tablenum{1}
\tablewidth{0pt}
\tablehead{
\colhead{Epoch} & \colhead{Rest frequency} & \multicolumn{2}{c}{Ring radius*} & \colhead{Fitting error} & \multicolumn{2}{c}{Converted coordinate (J2000)}& $\sqrt{\Delta\alpha^{2} + \Delta\delta^{2}}$ \\
&\colhead{(GHz)}  & \colhead{(mas)} & \colhead{(AU)} & \colhead{(mas)} & \colhead{R.A.} & \colhead{Dec.} & (mas)
}

\startdata
1& 43.122   & 13.35 &2.27&0.34 & 	 &\\
(May 21, 2015)& 86.243 & 12.72 &2.16&1.08 & 11:00:33.8204 &--18:19:29.619 & 4.1\\
\hline
2& 43.122   & 13.84 & 2.35 &0.20 & & \\
(Jan 7, 2016)& 86.243  & 11.76 & 2.00&0.73 & 11:00:33.8202 &--18:19:29.621 & 3.7 \\
\hline
3& 43.122  & 13.45 & 2.29 & 0.52 &   & \\
(Jan 26, 2016)& 86.243 & 11.96 &  2.03& 0.87 & 11:00:33.8201 &--18:19:29.616& 3.9\\
\enddata
\tablenotetext{*}{The reference coordinate of R Crt used in the observations is R.A.=11:00:33.850 Dec.=--18:19:29.60 (J2000).
A distance of 170 pc was assumed to calculate the ring diameter by AU unit \citep{2001PASJ...53.1231I}. Fitting error column  indicates RMS values of the fitting.}
\end{deluxetable}

\section{Discussion} 
\subsection{Comparison among the SiO v=1, 2 J=1$\rightarrow$0 (43.1, 42.8 GHz) and v=1, J=2$\rightarrow$1 (86.2 GHz) masers} \label{sec:Discussion1}

The most striking results from our observations is the discovery that the 86.2 GHz SiO maser spots are located closer to the central star, with a wider velocity range than those of the 43.1 and 42.8 GHz SiO masers, as shown in Figure 4 and Table 1. This is the opposite of what hitherto has been found. In the case of fundamental-pulsators such as WX Psc, R Leo, and $\chi$ Cyg \citep{2004A&amp;A...426..131S,2007A&amp;A...468L...1S} the 86.2 GHz maser was distributed in the comparable or more distant regions (up to $\sim$30\%) from the central star than the 43.1 and 42.8 GHz masers. Line overlap effects between the H$_{2}$O and SiO emission were proposed for the interpretation of 86.2 GHz maser location \citep{2014A&A...565A.127D}. Radiative Transfer (RT) models based on the large velocity gradient (LVG) assumption, which suppose a localized maser amplification, predict that 86.2 GHz maser would be stronger and located at the distant region from the central star compared to the 43.1 GHz maser with comparable velocity ranges \citep{2002A&amp;A...386..256H,2009MNRAS.394...51G}. Such results are consistent with the previous VLBI observations for the fundamental-pulsators \citep{2004A&amp;A...426..131S,2007A&amp;A...468L...1S}, but not for R Crt.

R Crt is a SRb type variable showing superimposed periodicities in the optical light curve and OH maser variability \citep{2001A&amp;A...378..522E,2003AcA....53..341P}. Our single-dish monitoring of H$_{2}$O and SiO masers also presents a signature of secondary variability (D. J. Kim et al. in prep.). SRb type variables are mainly overtone-pulsators, which have a short optical period with superposed periodicities, whereas other variables (SRa, SRc, and LPV) are characterised by a fundamental pulsation mode. The overtone-pulsator would result in more turbulent environment across the CSE, and it may produce different physical conditions for the masers compared to the fundamental-pulsators. In addition, \cite{1998A&A...334.1037H} detected high velocity components in the 86.2 GHz maser, which exceed the terminal velocity of the host AGB star. A statistical study points out that the high velocity wings of the 86.2 GHz maser lines dominantly appear in the SRb type variables \citep{2015AJ....149..100M}. This tendency is not seen in SRa, SRc, and long period variables.

On the other hand, the non-local radiative transfer (RT) model occasionally shows a slightly wider velocity range of the 86.2 GHz maser than the 43.1 GHz maser, supporting our observational results \citep{2012A&amp;A...545A.136Y,2015AJ....149..100M}. The non-local RT model considers all the velocity coherent regions along the line of sight to reflect the influence of distant maser clumps, which can contribute to the local maser amplification. This process is more likely to happen in overtone-pulsators rather than fundamental-pulsators, due to their complex dynamical environment induced by short and overlapping shock waves. Thus, the non-local RT model would be better at reproducing the 86.2 GHz SiO maser lines with a wider velocity range and maser distributions for SRb type variables than the RT model in fundamental-pulsators based on the LVG assumption \citep{2002A&amp;A...386..256H,2009MNRAS.394...51G}. 
  
However, the non-local RT model has never predicted that the 86.2 GHz maser features could appear at a smaller radius than those of the 43.1 GHz maser. The non-local RT model \citep{2012A&amp;A...545A.136Y} used the hydrodynamic solutions for a fundamental-pulsator. Therefore, further studies of the non-local RT model based on the hydrodynamic CSE model for overtone-pulsators are required for interpreting our results and for evaluating the effect of different pulsation modes on the various SiO maser features. In addition, we need to consider the high fractional missing flux of the 86.2 GHz maser (up to 80 \%) and relatively poor spatial sensitivity of the KVN, because is is possible that the 86.2 GHz maser features in the outer region of the 43.1 GHz maser of R Crt could have been resolved out or undetected due to their weak intensity. Therefore, following-up VLBI observations for R Crt and other SRb type variables are also required to confirm the inner distribution of the 86.2 GHz SiO masers in overtone-pulsators. Additionally adding more antennas and shorter baselines to the KVN would clarify these questions; such an enhancement is being planned.

\subsection{The stellar position and development of asymmetric structures in maser features} \label{sec:Discussion2}

The stellar position is a crucial parameter for analyzing the morphology and dynamics of the SiO and H$_{2}$O masers. Our astrometric observation scheme, using PR and SFPR, has provided the accurate positions of the SiO and H$_{2}$O maser spots. The position of the central star is estimated by a ring-fitting method to the SiO maser features. The fitting results and errors are listed in Table 1. The SFPR technique typically results in a relative positional error of less than 1 mas between the H$_{2}$O and SiO masers \citep{2018NatCo...9.2534Y}. This is dominated by the ring fitting error, which is about 1.87 mas corresponding to three times of the mean RMS value of the fittings. The total relative positional errors between the central star and the H$_{2}$O maser spots will be less than 3 mas, which is 8 times more accurate than the positional error of the central star determined by the three-dimensional velocity field of the 22.2 GHz H$_{2}$O maser \citep{2001PASJ...53.1231I}.
 
The astrometric performance of the PR observation using the KVN has not yet been explicitly demonstrated, as this aspect is still undergoing commissioning. However if we extrapolate from the expected behavior as found in \citet{2006A&A...452.1099P}, we obtain an astrometrical positional error, for a 3.06$^o$ calibrator-source separation, of about $\sim$2 mas. Combined with the ring fitting errors mentioned above we expect positional accuracies of $\sim$4 mas. In Table 1, we find positional differences of 2 mas (R.A.) and 5 mas (Dec.) between epochs 2 and 3, which must be due to measurement errors as there is only a short time separation of 19 days. This is consistant with the expected astrometric error. Therefore, we estimated the absolute stellar position (R.A.=11:00:33.8201, Dec.=--18:19:29.618) as the mean position obtained from epochs 2 and 3. The observed {\sc gaia} position for their reference epoch (Jun 2015), marked with a black cross in Figure 3, can be compared to our epoch 1 (May 2015) data directly as the proper motions and parallax contributions will be negligible. In this case the {\sc gaia} coordinates are 11:00:33.820623, --18:19:29.6219571, which are offset from our epoch 1 position by 3, 3 mas on the sky.

In Figure 1, the asymmetric spatial distribution of the H$_{2}$O maser with respect to the ring-like structure of the SiO maser cannot be interpreted with a spherical expanding shell structure. The observational results of the Japanese VLBI Network showed a possible bipolar outflow based on the three-dimensional velocity field of the H$_{2}$O maser \citep{2001PASJ...53.1231I}. However, the H$_{2}$O maser in Figure 1 shows a possible one-side outflow toward the southern part of the estimated position of the central star in Figure 3. Also, the H$_{2}$O maser in Figure 2 shows a significant intensity variation between May 2015 and Jan 2016. In the case of the red hypergiant star NML Cyg which shows a bipolar outflow in the H$_{2}$O maser feature, the central star is located close to the prominent blue-shifted H$_{2}$O maser features and not close to the center between the red-shifted and blue-shifted outflow features \citep{2012A&A...544A..42Z}. It is still uncertain what kind of physical process causes the development of the asymmetric structure of the H$_{2}$O maser. To investigate the development process of the highly asymmetric one-side outflow features of the H$_{2}$O maser from the ring-like SiO maser features in R Crt, we may need to measure the proper motion of both the H$_{2}$O and SiO masers through intensive VLBI monitoring observations.

\section{Summary} \label{sec:Summary}
Simultaneous VLBI monitoring observations of the H$_{2}$O 6$_{16}$--5$_{23}$ and SiO v=1, 2, J=1$\rightarrow$0 and v=1, J=2$\rightarrow$1, J=3$\rightarrow$2 masers toward the semiregular variable R Crt were performed with the KVN from Oct 2014 to Feb 2016.
We obtained high precision ``bonafide'' astrometrically registered multi-frequency maps of the 22.2 GHz H$_{2}$O and 43.1/42.8/86.2 GHz SiO masers at three epochs with the SFPR method. The SiO masers show a ring-like feature, while the H$_{2}$O masers show a very asymmetric one-sided outflow, which is located only at the southern part of the central star. Based on these astrometrically registered maps, we determined the position of the central star with an accuracy of 3 mas to the H$_{2}$O masers. The estimated stellar position is consistent with {\sc gaia} DR2 data. Furthermore, the SiO v=1, J=2$\rightarrow$1 maser spots are distributed in the inner-most region, with a 15 \% smaller radius, compared to those of the SiO v=1, J=1$\rightarrow$0 maser. Some maser spots of the SiO v=1, J=2$\rightarrow$1 maser also show highly blue-shifted components which exceed the terminal velocity of R Crt. We suggest that these features may be related with the characteristics of the overtone pulsation mode of the SRb-type R Crt associated with complex dynamics in its CSE. However, we need to investigate other overtone-pulsators to confirm whether those properties are common.

\acknowledgments
This work was supported by the Basic and Fusion Research Programs (2014-2017). We are grateful to all of the staff members at KVN who helped to operate the array and the single dish telescope and to correlate the data. The KVN is a facility operated by KASI (Korea Astronomy and Space Science Institute), which is under the protection of the National Research Council of Science and Technology (NST). The KVN operations are supported by KREONET (Korea Research Environment Open NETwork) which is managed and operated by KISTI (Korea Institute of Science and Technology Information).
\bibliographystyle{aasjournal}

\begin{thebibliography}{}
\expandafter\ifx\csname natexlab\endcsname\relax\def\natexlab#1{#1}\fi
\providecommand{\url}[1]{\href{#1}{#1}}
\providecommand{\dodoi}[1]{doi:~\href{http://doi.org/#1}{\nolinkurl{#1}}}
\providecommand{\doeprint}[1]{\href{http://ascl.net/#1}{\nolinkurl{http://ascl.net/#1}}}
\providecommand{\doarXiv}[1]{\href{https://arxiv.org/abs/#1}{\nolinkurl{https://arxiv.org/abs/#1}}}

\bibitem[{{Cho} {et~al.}(2018){Cho}, {Yun}, {Kim}, {Yoon}, {Kim}, {Choi},
  {Dodson}, {Rioja}, \& {Imai}}]{2018IAUS..336..359C}
{Cho}, S.-H., {Yun}, Y., {Kim}, J., {et~al.} 2018, in IAU Symposium, Vol. 336,
  Astrophysical Masers: Unlocking the Mysteries of the Universe, ed.
  A.~{Tarchi}, M.~J. {Reid}, \& P.~{Castangia}, 359--364

\bibitem[{{Desmurs} {et~al.}(2014){Desmurs}, {Bujarrabal}, {Lindqvist},
  {Alcolea}, {Soria-Ruiz}, \& {Bergman}}]{2014A&A...565A.127D}
{Desmurs}, J.-F., {Bujarrabal}, V., {Lindqvist}, M., {et~al.} 2014, \aap, 565,
  A127, \dodoi{10.1051/0004-6361/201423550}

\bibitem[{{Diamond} \& {Kemball}(2003)}]{2003ApJ...599.1372D}
{Diamond}, P.~J., \& {Kemball}, A.~J. 2003, \apj, 599, 1372,
  \dodoi{10.1086/379347}

\bibitem[{{Diamond} {et~al.}(1994){Diamond}, {Kemball}, {Junor}, {Zensus},
  {Benson}, \& {Dhawan}}]{1994ApJ...430L..61D}
{Diamond}, P.~J., {Kemball}, A.~J., {Junor}, W., {et~al.} 1994, \apjl, 430,
  L61, \dodoi{10.1086/187438}

\bibitem[{{Dodson} {et~al.}(2014){Dodson}, {Rioja}, {Jung}, {Sohn}, {Byun},
  {Cho}, {Lee}, {Kim}, {Kim}, {Oh}, {Han}, {Je}, {Chung}, {Wi}, {Kang}, {Lee},
  {Chung}, {Kim}, {Kim}, {Lee}, {Roh}, {Oh}, {Yeom}, {Song}, \&
  {Kang}}]{2014AJ....148...97D}
{Dodson}, R., {Rioja}, M.~J., {Jung}, T.-H., {et~al.} 2014, \aj, 148, 97,
  \dodoi{10.1088/0004-6256/148/5/97}

\bibitem[{{Etoka} {et~al.}(2001){Etoka}, {B{\l}aszkiewicz}, {Szymczak}, \& {Le
  Squeren}}]{2001A&amp;A...378..522E}
{Etoka}, S., {B{\l}aszkiewicz}, L., {Szymczak}, M., \& {Le Squeren}, A.~M.
  2001, \aap, 378, 522, \dodoi{10.1051/0004-6361:20011184}

\bibitem[{{Gray} {et~al.}(2009){Gray}, {Wittkowski}, {Scholz}, {Humphreys},
  {Ohnaka}, \& {Boboltz}}]{2009MNRAS.394...51G}
{Gray}, M.~D., {Wittkowski}, M., {Scholz}, M., {et~al.} 2009, \mnras, 394, 51,
  \dodoi{10.1111/j.1365-2966.2008.14237.x}

\bibitem[{{Han} {et~al.}(2008){Han}, {Lee}, {Kang}, {Je}, {Chung}, {Wi},
  {Sasao}, \& {Wylde}}]{2008IJIMW..29...69H}
{Han}, S.-T., {Lee}, J.-W., {Kang}, J., {et~al.} 2008, International Journal of
  Infrared and Millimeter Waves, 29, 69, \dodoi{10.1007/s10762-007-9296-7}

\bibitem[{{Herpin} {et~al.}(1998){Herpin}, {Baudry}, {Alcolea}, \&
  {Cernicharo}}]{1998A&A...334.1037H}
{Herpin}, F., {Baudry}, A., {Alcolea}, J., \& {Cernicharo}, J. 1998, \aap, 334,
  1037

\bibitem[{{Humphreys} {et~al.}(2002){Humphreys}, {Gray}, {Yates}, {Field},
  {Bowen}, \& {Diamond}}]{2002A&amp;A...386..256H}
{Humphreys}, E.~M.~L., {Gray}, M.~D., {Yates}, J.~A., {et~al.} 2002, \aap, 386,
  256, \dodoi{10.1051/0004-6361:20020202}

\bibitem[{{Imai} {et~al.}(2003){Imai}, {Shibata}, {Marvel}, {Diamond}, {Sasao},
  {Miyoshi}, {Inoue}, {Migenes}, \& {Murata}}]{2003ApJ...590..460I}
{Imai}, H., {Shibata}, K.~M., {Marvel}, K.~B., {et~al.} 2003, \apj, 590, 460,
  \dodoi{10.1086/374887}

\bibitem[{{Ishitsuka} {et~al.}(2001){Ishitsuka}, {Imai}, {Omodaka}, {Ueno},
  {Kameya}, {Sasao}, {Morimoto}, {Miyaji}, {Nakajima}, \&
  {Watanabe}}]{2001PASJ...53.1231I}
{Ishitsuka}, J.~K., {Imai}, H., {Omodaka}, T., {et~al.} 2001, \pasj, 53, 1231,
  \dodoi{10.1093/pasj/53.6.1231}

\bibitem[{{Kahane} \& {Jura}(1994)}]{1994A&A...290..183K}
{Kahane}, C., \& {Jura}, M. 1994, \aap, 290, 183

\bibitem[{{Kim} {et~al.}(2010){Kim}, {Cho}, {Oh}, \&
  {Byun}}]{2010ApJS..188..209K}
{Kim}, J., {Cho}, S.-H., {Oh}, C.~S., \& {Byun}, D.-Y. 2010, \apjs, 188, 209,
  \dodoi{10.1088/0067-0049/188/1/209}

\bibitem[{{Knapp} {et~al.}(1998){Knapp}, {Young}, {Lee}, \&
  {Jorissen}}]{1998ApJS..117..209K}
{Knapp}, G.~R., {Young}, K., {Lee}, E., \& {Jorissen}, A. 1998, \apjs, 117,
  209, \dodoi{10.1086/313111}

\bibitem[{{Lee} {et~al.}(2011){Lee}, {Byun}, {Oh}, {Han}, {Je}, {Kim}, {Wi},
  {Cho}, {Sohn}, {Kim}, {Lee}, {Oh}, {Song}, {Kang}, {Chung}, {Lee}, {Oh},
  {Bae}, {Yun}, {Lee}, {Kim}, {Chung}, {Roh}, {Lee}, {Kim}, {Ryoung Kim},
  {Yeom}, {Kurayama}, {Jung}, {Park}, {Kim}, {Yoon}, \&
  {Kim}}]{2011PASP..123.1398L}
{Lee}, S.-S., {Byun}, D.-Y., {Oh}, C.~S., {et~al.} 2011, \pasp, 123, 1398,
  \dodoi{10.1086/663326}

\bibitem[{{Ma} {et~al.}(2009){Ma}, {Arias}, {Bianco}, {Boboltz}, {Bolotin},
  {Charlot}, {Engelhardt}, {Fey}, {Gaume}, {Gontier}, {Heinkelmann}, {Jacobs},
  {Kurdubov}, {Lambert}, {Malkin}, {Nothnagel}, {Petrov}, {Skurikhina},
  {Sokolova}, {Souchay}, {Sovers}, {Tesmer}, {Titov}, {Wang}, {Zharov},
  {Barache}, {Boeckmann}, {Collioud}, {Gipson}, {Gordon}, {Lytvyn},
  {MacMillan}, \& {Ojha}}]{2009ITN....35....1M}
{Ma}, C., {Arias}, E.~F., {Bianco}, G., {et~al.} 2009, IERS Technical Note, 35,
  1

\bibitem[{{McIntosh} \& {Indermuehle}(2015)}]{2015AJ....149..100M}
{McIntosh}, G., \& {Indermuehle}, B. 2015, \aj, 149, 100,
  \dodoi{10.1088/0004-6256/149/3/100}

\bibitem[{{Nyman} {et~al.}(1992){Nyman}, {Booth}, {Carlstrom}, {Habing},
  {Heske}, {Sahai}, {Stark}, {van der Veen}, \&
  {Winnberg}}]{1992A&AS...93..121N}
{Nyman}, L.-A., {Booth}, R.~S., {Carlstrom}, U., {et~al.} 1992, \aaps, 93, 121

\bibitem[{{Olofsson} {et~al.}(2002){Olofsson}, {Gonz{\'a}lez Delgado},
  {Kerschbaum}, \& {Sch{\"o}ier}}]{2002A&A...391.1053O}
{Olofsson}, H., {Gonz{\'a}lez Delgado}, D., {Kerschbaum}, F., \& {Sch{\"o}ier},
  F.~L. 2002, \aap, 391, 1053, \dodoi{10.1051/0004-6361:20020841}

\bibitem[{{Perrin} {et~al.}(2015){Perrin}, {Cotton}, {Millan-Gabet}, \&
  {Mennesson}}]{2015A&amp;A...576A..70P}
{Perrin}, G., {Cotton}, W.~D., {Millan-Gabet}, R., \& {Mennesson}, B. 2015,
  \aap, 576, A70, \dodoi{10.1051/0004-6361/201425110}

\bibitem[{{Pojmanski}(2003)}]{2003AcA....53..341P}
{Pojmanski}, G. 2003, \actaa, 53, 341

\bibitem[{{Pradel} {et~al.}(2006){Pradel}, {Charlot}, \&
  {Lestrade}}]{2006A&A...452.1099P}
{Pradel}, N., {Charlot}, P., \& {Lestrade}, J.-F. 2006, \aap, 452, 1099,
  \dodoi{10.1051/0004-6361:20053021}

\bibitem[{{Reid} \& {Menten}(1990)}]{1990ApJ...360L..51R}
{Reid}, M.~J., \& {Menten}, K.~M. 1990, \apjl, 360, L51, \dodoi{10.1086/185810}

\bibitem[{{Reid} \& {Menten}(2007)}]{2007ApJ...671.2068R}
---. 2007, \apj, 671, 2068, \dodoi{10.1086/523085}

\bibitem[{{Reid} \& {Moran}(1981)}]{1981ARA&amp;A..19..231R}
{Reid}, M.~J., \& {Moran}, J.~M. 1981, \araa, 19, 231,
  \dodoi{10.1146/annurev.aa.19.090181.001311}

\bibitem[{{Rioja} \& {Dodson}(2011)}]{2011AJ....141..114R}
{Rioja}, M., \& {Dodson}, R. 2011, \aj, 141, 114,
  \dodoi{10.1088/0004-6256/141/4/114}

\bibitem[{{Rosen} {et~al.}(1978){Rosen}, {Moran}, {Reid}, {Walker}, {Burke},
  {Johnston}, \& {Spencer}}]{1978ApJ...222..132R}
{Rosen}, B.~R., {Moran}, J.~M., {Reid}, M.~J., {et~al.} 1978, \apj, 222, 132,
  \dodoi{10.1086/156129}

\bibitem[{{Samus'} {et~al.}(2017){Samus'}, {Kazarovets}, {Durlevich},
  {Kireeva}, \& {Pastukhova}}]{2017ARep...61...80S}
{Samus'}, N.~N., {Kazarovets}, E.~V., {Durlevich}, O.~V., {Kireeva}, N.~N., \&
  {Pastukhova}, E.~N. 2017, Astronomy Reports, 61, 80,
  \dodoi{10.1134/S1063772917010085}

\bibitem[{{Soria-Ruiz} {et~al.}(2007){Soria-Ruiz}, {Alcolea}, {Colomer},
  {Bujarrabal}, \& {Desmurs}}]{2007A&amp;A...468L...1S}
{Soria-Ruiz}, R., {Alcolea}, J., {Colomer}, F., {Bujarrabal}, V., \& {Desmurs},
  J.-F. 2007, \aap, 468, L1, \dodoi{10.1051/0004-6361:20077554}

\bibitem[{{Soria-Ruiz} {et~al.}(2004){Soria-Ruiz}, {Alcolea}, {Colomer},
  {Bujarrabal}, {Desmurs}, {Marvel}, \& {Diamond}}]{2004A&amp;A...426..131S}
{Soria-Ruiz}, R., {Alcolea}, J., {Colomer}, F., {et~al.} 2004, \aap, 426, 131,
  \dodoi{10.1051/0004-6361:20041139}

\bibitem[{{Szymczak} \& {Le Squeren}(1999)}]{1999MNRAS.304..415S}
{Szymczak}, M., \& {Le Squeren}, A.~M. 1999, \mnras, 304, 415,
  \dodoi{10.1046/j.1365-8711.1999.02321.x}

\bibitem[{{Yoon} {et~al.}(2018){Yoon}, {Cho}, {Yun}, {Choi}, {Dodson}, {Rioja},
  {Kim}, {Imai}, {Kim}, {Yang}, \& {Byun}}]{2018NatCo...9.2534Y}
{Yoon}, D.-H., {Cho}, S.-H., {Yun}, Y., {et~al.} 2018, Nature Communications,
  9, 2534, \dodoi{10.1038/s41467-018-04767-8}

\bibitem[{{Yun} \& {Park}(2012)}]{2012A&amp;A...545A.136Y}
{Yun}, Y.~J., \& {Park}, Y.-S. 2012, \aap, 545, A136,
  \dodoi{10.1051/0004-6361/201218953}

\bibitem[{{Zhang} {et~al.}(2012){Zhang}, {Reid}, {Menten}, {Zheng}, \&
  {Brunthaler}}]{2012A&A...544A..42Z}
{Zhang}, B., {Reid}, M.~J., {Menten}, K.~M., {Zheng}, X.~W., \& {Brunthaler},
  A. 2012, \aap, 544, A42, \dodoi{10.1051/0004-6361/201219587}

\end{thebibliography}

\end{document}